%% file: main.tex
\documentclass[]{fairmeta}

\usepackage[utf8]{inputenc} 
\usepackage[T1]{fontenc}    
\usepackage{hyperref}       
\usepackage{url}            
\usepackage{booktabs}       
\usepackage{amsfonts}       
\usepackage{nicefrac}       
\usepackage{microtype}      
\usepackage{xcolor}         
\usepackage{multirow}
\usepackage{graphicx}
\usepackage{makecell}
\usepackage{subcaption}
\usepackage{enumitem}
\usepackage{xspace}

\newcolumntype{L}[1]{>{\raggedright\let\newline\\\arraybackslash\hspace{0pt}}m{#1}}
\newcolumntype{C}[1]{>{\centering\let\newline\\\arraybackslash\hspace{0pt}}m{#1}}
\newcolumntype{R}[1]{>{\raggedleft\let\newline\\\arraybackslash\hspace{0pt}}m{#1}}
\makeatletter 
\newcommand{\thickhline}{%
    \noalign {\ifnum 0=`}\fi \hrule height 1pt
    \futurelet \reserved@a \@xhline
}
\makeatother

\usepackage{listings}
\usepackage{xcolor}
\usepackage{float}
\newfloat{lstfloat}{htbp}{lop}
\floatname{lstfloat}{Listing}
\definecolor{commentsColor}{rgb}{0.497495, 0.497587, 0.497464}
\definecolor{keywordsColor}{rgb}{0.000000, 0.000000, 0.635294}
\definecolor{stringColor}{rgb}{0.558215, 0.000000, 0.135316}
\lstdefinestyle{mystyle}{
    basicstyle=\footnotesize\ttfamily,
    captionpos=b,
    breaklines=true,
    breakindent=0.5em,
    tabsize=2,
    frame=b,
    showstringspaces=false,
    numberstyle=\tiny\color{commentsColor},
    rulecolor=\color{black},
    commentstyle=\color{commentsColor}\textit,
    stringstyle=\color{stringColor},
    keywordstyle=\color{keywordsColor},
    emphstyle=\color{keywordsColor},
    escapeinside={(*@}{@*)},
}
\lstdefinestyle{tinyinline}{
    basicstyle=\scriptsize\ttfamily,
    captionpos=b,
    breaklines=true,
    breakindent=0.5em,
    tabsize=2,
    frame=b,
    showstringspaces=false,
    numberstyle=\tiny\color{commentsColor},
    rulecolor=\color{black},
    commentstyle=\color{commentsColor}\textit,
    stringstyle=\color{stringColor},
    keywordstyle=\color{keywordsColor},
    emphstyle=\color{keywordsColor},
    escapeinside={(*@}{@*)},
}
\AtBeginDocument{\DeclareCaptionSubType{lstlisting}}

\title{Finding Missed Code Size Optimizations\\*in Compilers using LLMs}
\author[1]{Davide Italiano}
\author[2]{Chris Cummins}

\affiliation[1]{Meta}
\affiliation[2]{FAIR at Meta}

\abstract{\input{abstract}}

\date{\today}
\correspondence{Davide Italiano at \email{davidino@meta.com}}

\begin{document}   

\maketitle

\input{sections/introduction}
\input{sections/methodology}
\input{sections/results}
\input{sections/related-work}
\input{sections/conclusions}

\bibliography{references}

\end{document}

%% file: abstract.tex
Compilers are complex, and significant effort has been expended on testing them.
Techniques such as random program generation and differential testing have proved highly effective and have uncovered thousands of bugs in production compilers.
The majority of effort has been expended on validating that a compiler produces \emph{correct} code for a given input, while less attention has been paid to ensuring that the compiler produces \emph{performant} code.

In this work we adapt differential testing to the task of identifying missed optimization opportunities in compilers.
We develop a novel testing approach which combines large language models (LLMs) with a series of differential testing strategies and use them to find missing code size optimizations in C / C++ compilers.

The advantage of our approach is its simplicity.
We offload the complex task of generating random code to an off-the-shelf LLM, and use heuristics and analyses to identify anomalous compiler behavior.
Our approach requires fewer than 150 lines of code to implement.
This simplicity makes it extensible.
By simply changing the target compiler and initial LLM prompt we port the approach from C / C++ to Rust and Swift, finding bugs in both.
To date we have reported 24 confirmed bugs in production compilers, and conclude that LLM-assisted testing is a promising avenue for detecting optimization bugs in real world compilers.

%% file: sections/introduction.tex
\section{Introduction}

\begin{figure*}
\begin{center}
\centerline{\includegraphics[width=.9\textwidth]{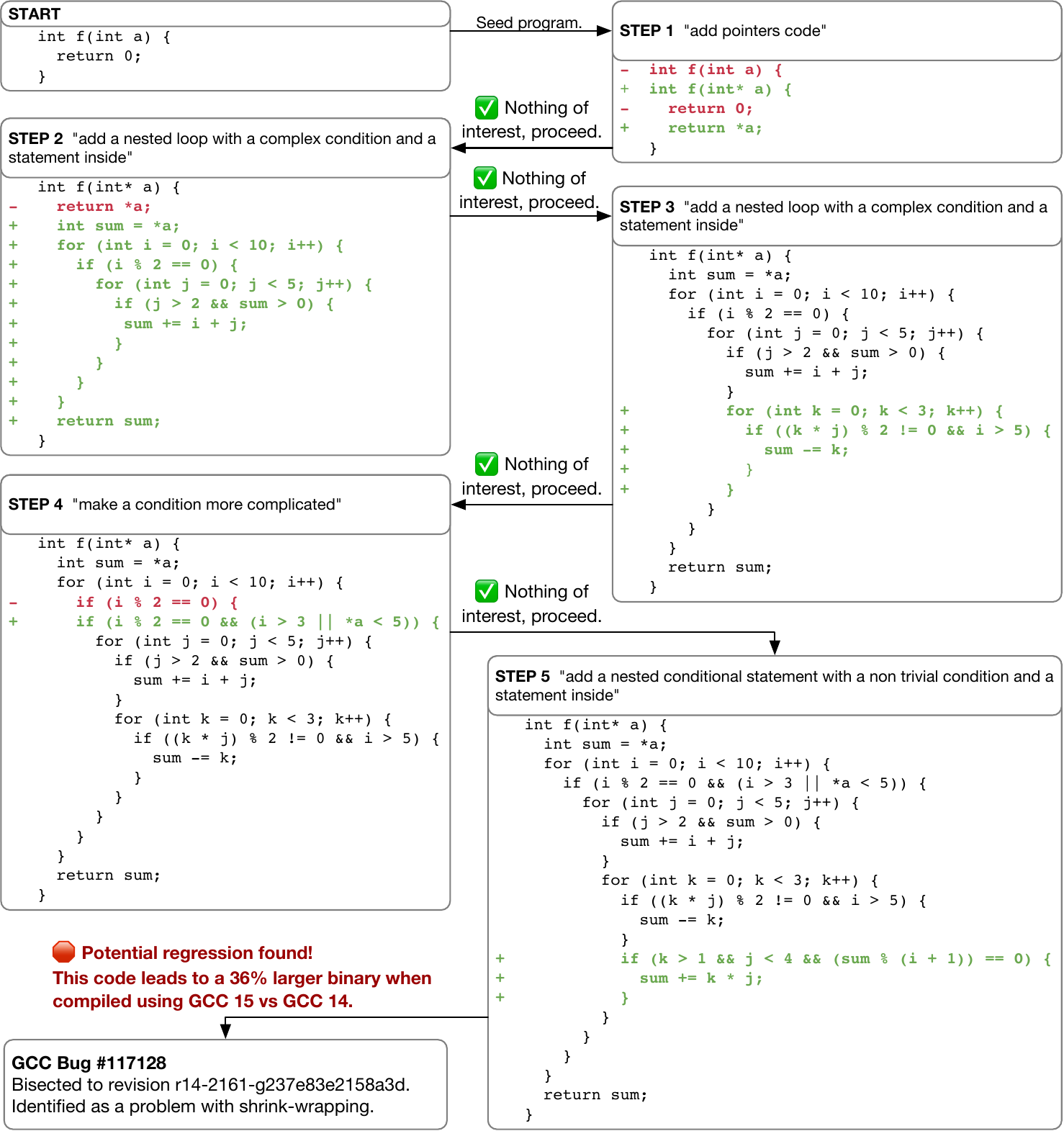}}
\caption{%
    An example of our technique. We instruct an LLM to incrementally mutate a program by randomly sampling a predetermined list of instructions. At each mutation step, an automatic differential testing strategy is used to detect missed optimizations. For this particular example one minute of compute was used and a 36\% code size regression was discovered.
}
\label{fig:motivation}
\end{center}
\end{figure*}

Significant effort has been put into testing and fuzzing compilers~\citep{chen2020survey,tang2020compiler}.
A common way to test compilers is to generate random programs that get fed into a compiler, using techniques such as differential testing to validate correctness of the generated binaries~\citep{difftest}.
Generating random programs is challenging because of the complex nature of code. Random program generators must be written for each language and maintained as the language evolves with new features.
This, in itself, requires a complex piece of software which requires extensive compiler and programming language expertise to develop.
For example, CSmith~\citep{csmith}, a highly effective random program generator which has been used to identify hundreds of bugs in C compilers, comprises well over 40,000 lines of handwritten code, and requires constantly updating as the language evolves.
Such an approach can prove prohibitively expensive for new programming languages, and can limit the effectiveness of testing even popular languages.

Additionally, test cases generated by random program generators are often large and hard to interpret, requiring an additional program reduction stage to make them useful in reporting bugs to compiler developers.
This requires further compute and language-specific tooling to be built, for example, using C-Reduce in the case of C programs~\citep{creduce}.
As an alternative to generating new programs, mutation testing takes as input a \emph{seed} code and modifies it, such as by mutating the parts of the code which are not executed over a given set of inputs~\citep{emi}.
As with random program generation, developing such tools requires a deep understanding of the target programming language features and static and dynamic analyses.

Prior to the advent of LLMs, there was a strand of research that attempted to leverage machine learning for test case generation~\citep{deepsmith,deepfuzz,learn-and-fuzz}.
The idea was to substitute the rule- and grammar-based test case generators with a learned generative model that could be stochastically sampled to produce new code snippets.
The main advantage of such an approach is the enormous reduction in human effort required to train a model vs build a random program generator.
Early work demonstrated some promise in generating plausible and interpretable test cases using deep recurrent neural networks, but the techniques struggled under the load of models that demonstrated a poor grasp of programming language syntax and semantics.
For example, in~\citep{deepsmith}, the Long Short-Term Memory network trained on OpenCL required on average 20 attempts to generate a single compilable code snippet.

With LLMs, the capabilities of models to generate and reason about code has improved markedly and there is now a plethora of research directions applying LLMs to different software domains~\citep{llm-swe-survey}.
While nascent, LLMs have already been used to fuzz the correctness of deep learning libraries~\citep{titanfuzz,fuzzgpt}, and C/C++ compilers~\citep{fuzz4all}.

While there is a great number of traditional compiler test case generators, and while machine learning and LLMs are being used to simplify their implementation, the automatic generation of compiler test cases to find missed code size optimizations has received little attention.
Yet, code size optimization is critical for embedded computing, mobile applications, and firmware software, and there is active research into novel code optimizations~\citep{hoag2024reordering,lee2024optimistic,lee2022efficient}.
In this work, we develop novel differential testing methodologies for the express purpose of discovering missed code size optimization opportunities.
Our contributions are as follows:

\begin{itemize}
  \item We present a novel mutation testing methodology which uses large language models to iteratively modify a starting code seed.
  \item We develop four differential testing strategies for finding missed code size optimizations in compilers.
  \item We implement our approach in fewer than 150 lines of code and use it to identify 24 bugs in production compilers across C/C++, Rust, and Swift. We release this tool open source.
\end{itemize}

%% file: sections/methodology.tex
\section{Methodology}

\begin{figure*}[t]
\begin{center}
\centerline{\includegraphics[width=.95\textwidth]{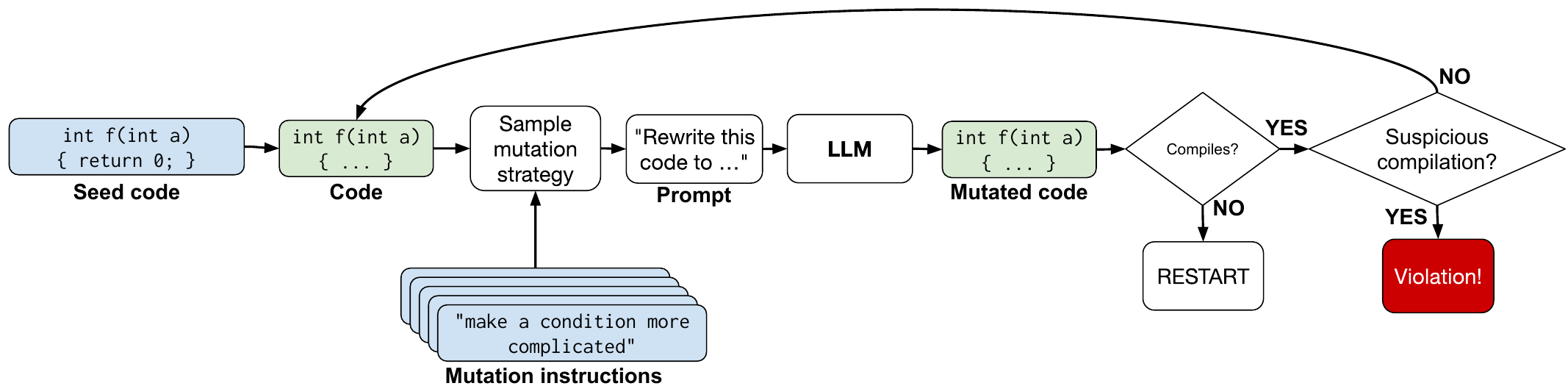}}
\caption{%
    Workflow of the automated testing methodology. The system takes two inputs provided by the user: a seed code and a list of mutation instructions (Section~\ref{sec:mutation}). Execution iterates until the code mutated by the LLM no longer compiles, or until a series of differential tests and analyses detect a suspicious compilation and trigger a violation (Section~\ref{sec:detecting-suspicious-compilations}).
}
\label{fig:flow}
\end{center}
\end{figure*}

In this section we describe how we identify missed code size optimization opportunities in compilers using LLMs.

Figure~\ref{fig:motivation} shows a demonstration of our approach.
Starting with a simple seed code, we iteratively instruct an LLM to mutate the code by randomly sampling from a preselected list of mutation instructions.
After each mutation, we compile the resulting code and apply a series of differential testing strategies to identify suspicious compilers.
Static and dynamic analyses are used to mitigate false positives, and once identified, suspicious compilation results are reported to the user.
In this case, after 5 mutations the system identified a 36\% code size regression between GCC 15 and GCC 14.

Figure~\ref{fig:flow} illustrates the workflow of the automated testing methodology. Our approach has two component stages: a method for mutating code, and a series of differential testing strategies to identify potential missed optimization bugs. We describe each in turn, followed by techniques for identifying false positives and detecting duplicate issues.

\subsection{Mutating code using LLMs}
\label{sec:mutation}

\begin{table}

\caption{Instructions used to mutate code. At each step we sample uniformly from this list and generate a prompt which we feed to an LLM.}
\centering
\begin{tabular}{L{.85\columnwidth}}
\toprule
\textbf{Control flow}\\
\midrule
``add a conditional statement with a statement inside''\\
``add a nested conditional statement with a non trivial condition and a statement inside''\\
``add a dead conditional statement with a statement inside''\\
``add a dead nested conditional statement with a non trivial condition and a statement inside''\\
``add a loop with a complex condition and statement inside''\\
``add a dead loop with a complex condition and statement inside''\\
``add a nested loop  with a complex condition and a statement inside''\\
``add a dead nested loop  with a complex condition and a statement inside''\\
\midrule
\textbf{Conditionals}\\
\midrule
``make a condition more complicated''\\
``make a dead condition more complicated''\\
\midrule
\textbf{Aggregates/pointers}\\
\midrule
``add array code''\\
``add pointers code''\\
``add struct code usage''\\
``add union code usage''\\
\midrule
\textbf{Function arguments}\\
\midrule
``add function arguments to a function that already\\*\hspace{.5cm}exists, no default arguments''\\
\bottomrule
\label{tab:actions}
\end{tabular}
\end{table}

Typical approaches to compiler test case generation requires defining a grammar of the target programming language, and then probabilistically sampling from this grammar, combined with rigorous static and dynamic analyses so as to generate new code which is both syntactically and semantically correct~\citep{yarpgen,csmith}. Such an approach guarantees that generated test cases are free from undefined behavior, but at the expense of complex generation logic. For example, CSmith~\citep{csmith} comprises over 40,000 lines of handwritten code.

We take a different approach. By forfeiting the correct-by-construction guarantee of a grammar-based generator, we are able to use a much simpler engine to generate code for testing. We start with a trivial input program snippet and use an off-the-shelf LLM to rewrite it in such a way as to incrementally add complexity. LLMs make syntactic and semantic errors, and this can make it more challenging to determine if anomalous compiler behavior is indicative of a true bug or a result of a mistake made by in the LLM output.
We accommodate for this in two ways: first, because we target missing optimization opportunities, we can permit a greater class of programs than functional tests, and second, by using heuristics and analyses to detect false positives, described in Section~\ref{sec:false-positives}.

\paragraph{Seed program}

As in prior mutation-based testing approaches, we start with a seed program. Because our iterative approach accumulates mutations, our seed programs can be very simple, shown in Listing~\ref{fig:seeds}.

\begin{figure}
\setcaptiontype{lstlisting}
\begin{subfigure}{.95\columnwidth}
\begin{lstlisting}
int f(int a) { return 0; }
\end{lstlisting}
\caption{C / C++}
\end{subfigure}
\begin{subfigure}{.95\columnwidth}
\begin{lstlisting}
#![no_main]
#[no_mangle]
pub fn f(a: i32) -> i32 { 0 }
\end{lstlisting}
\caption{Rust}
\end{subfigure}
\begin{subfigure}{.95\columnwidth}
\begin{lstlisting}
func f(a: Int) -> Int { return a }
\end{lstlisting}
\caption{Swift}
\end{subfigure}
\caption{Seed programs for different programming languages, used as the starting point for mutation. In all three languages the seed code contains a single empty function with an integer argument. From this, the LLM incrementally expands the scope and complexity of the code, directed by our mutation prompts.}
\label{fig:seeds}
\end{figure}

\paragraph{Mutation prompts}

At each step of the iterative testing process we build a prompt that instructs the model to mutate the code by randomly sampling from a list of predetermined instructions. We then assemble the instruction and current code state into a prompt using the template shown in Listing~\ref{lst:prompt-template}, and feed it to an off-the-shelf LLM, which responds with mutated code.

In this work we use the freely available Llama 3.1 models~\citep{llama3}. These models have been pre-trained and instruction fine tuned on vast corpora of data, with a ``knowledge cutoff'' of December 2023. For language features added late, one would need a more recently trained LLM.

\begin{figure}
\setcaptiontype{lstlisting}
\begin{lstlisting}
Given the following ${language} program,
please ${instruction}:

${code}
\end{lstlisting}
\caption{Template used to generate LLM prompts.}
\label{lst:prompt-template}
\end{figure}

\paragraph{Mutation instructions}

We sample uniformly from a list of predetermined mutation instructions, shown in Table~\ref{tab:actions}. We curated several different kind of instructions, that can be divided into four logical categories:

\textit{Control flow modifications}: we instruct the model to inject some new control flow into the code. This comes in two form: loops or conditional controls statements (\lstinline{if} statements). Additionally we control the nesting factor asking the model to generate either a top-level structure or a nested structure inside existing code.

\textit{Aggregate mutations}: we instruct the model to add code that contains aggregate structures. In our code we take in consideration 4 classes of aggregates: arrays, unions, structures and classes. For languages that don't support C-style unions, e.g. Swift, we replace them with enumerations.

\textit{Pointer code}: We instruct the model to add code that contains pointers to variables defined in the code.

\textit{Mutation of existing conditionals}: if \lstinline{for} loops and \lstinline{if} statements are already present in the program, we instruct the model to modify the existing condition to make it more complicated without changing the semantics. For example, a transformation that the model has made in response to this prompt is: \lstinline!if (x == 10) {...}! $\rightarrow$
\lstinline!if (x >= 10 && x <= 10) {...}!.

Many of the code mutation instructions come in two flavors: \textit{dead code} and \textit{live code}, where dead code is a mutation that changes code that is not executed as part of the regular flow of the program.

The list of mutation instructions is user configurable. We chose these instructions to represent a diverse set of program transformations, but they are not comprehensive, and further could be adapted for domain specific uses if needed.

\subsection{Differential testing strategies for discovering missed optimizations}
\label{sec:detecting-suspicious-compilations}

In this section we describe four strategies for differential testing that we employ to identify potential missed optimization opportunities. Once a strategy is selected it is kept constant for the duration of an iterative mutation testing session.

\subsubsection{Dead code differential testing}
\label{sec:dead-code-difftest}

The idea with dead code differential testing is that since our initial seed code evaluates to a constant, the addition of dead code should not change the semantics of the program, so the code generated should be the same. To perform this type of differential test, we disable all mutation instructions except for those which instruct the LLM to insert or modify dead code. The idea is that of telling the LLM to keep adding code and then recompile and compare the code that gets generated. If the code generated changes, it means the compiler failed to prove that the code was dead, pessimizing. This differential testing strategy is particularly interesting because it doesn't need another compiler to be tested against. One can just build the same program twice with the same compiler and compare the code that gets generated.

\subsubsection{Optimization pipeline differential testing}
\label{sec:pipeline-difftest}

The second differential testing strategy is based on the intuition that certain optimization pipelines try to minimize a metric at the expense of everything else. For example, there is an optimization pipeline that tries to minimize code size at every cost, even at the expense of performance. For LLVM and GCC this is enabled using \lstinline{-Oz}, and for Rust using \lstinline{opt-level=z}. We exploit this for differential testing by comparing the code generated by the same compiler using different optimization pipelines (e.g. \lstinline{-O3}), and triggers a violation if the size of the code at an optimization level different from \lstinline{-Oz} is smaller from the one at \lstinline{-Oz}, multiplied by some configurable sensitivity threshold. We found a sensitivity threshold of 5\% to be effective.

\subsubsection{Single-compiler differential testing}
\label{sec:single-compiler-difftest}

The third differential testing strategy relies on comparing the code generated between different versions of the same compiler. This is useful to find size optimization progressions and regressions. We compare already-released versions of a compiler against the bleeding edge version (trunk/nightly, depending on the language). If the code size regresses, a violation is triggered - a potential candidate for reporting a problem.

\subsubsection{Multi-compiler differential testing}
\label{sec:multi-compiler-difftest}

The forth strategy is a ``basic'' differential testing methodology. We compare, when available, multiple different compilers to compare the code that gets generated and if the sizes are sufficiently different, we can flag the one that has larger size as a violation and potential indicator that this compiler misses and optimization found by the other compiler. We use a threshold of 10\% to establish ``sufficiently different''.

\subsection{Detecting false positives}
\label{sec:false-positives}

The limitation of using an LLM to write code instead of a grammar-based generator that has full control is that there's the risk of introducing bad examples.
While we are unable to fully eliminate the possibility of incorrect code generated by an LLM, we found that we can sufficiently negate the risk using heuristics and validation tools such that we have yet to encounter a false positive that has not been detected by the three techniques described here:

\emph{Dead code detection.} Of course, when we instruct the LLM to insert dead code, we can't guarantee that the LLM will obey the instruction, and we need a way to verify that the mutated code truly contains only dead code. The problem in general is undecidable, but we can use a proxy metric. Since we control the function signature of the mutated code (e.g. \lstinline{f()} for C/C++), we can compile an executable program from the test code by inserting a main method that calls it. We run the generated program with counter instrumentation (for coverage) and we wait until the program terminates (if it does), then compare the inserted code and make sure the counter values are identical. If the LLM corrupts the function signature, compilation will fail and the process will restart. There are some limitations to this approach -- debug info mapping isn't always reliable, so we might end up mapping from an instruction in assembly to the wrong line, causing false positives and negatives, or the program might never terminate. Nevertheless, we did find these to be effective for our use case, and have yet to encounter a real false positive using this strategy.

\emph{Sanitization}: Every test case generated is compiled with sanitizers to rule out some classes of potential memory violation problems and undefined behavior. For C/C++, we use UBSan and ASan from LLVM~\citep{llvm}.
To reduce the set of false positives further, when available, we run the examples through Compcert~\citep{compcert} that in its "interpreter" mode finds undefined behavior.
In our experiments we found that these sanitization steps rejected less than 10\% of candidates.

\emph{Monotonically increasing size}: This is a static heuristic, and it's the easiest one to verify of the all three, so it's the first one we end up employing. It has a higher rate of false negatives. The idea behind it is that in some cases of undefined behavior, the compiler is free to label code as unreachable, and remove. Since our iterative mutation testing incrementally adds complexity for programs with undefined behavior, the optimizer tends to ``remove code'', rather than adding new code. Therefore, code size can be used as a proxy metric for complexity.

\subsection{Detecting duplicates}

As in all undirected test case generators for compilers, our approach may yield numerous unique test cases that trigger violations, but that are duplicates of the same underlying bug. We employ two techniques to help mitigate this.

\emph{Release screening}: First, we look at the code and we compile with publicly available releases of compilers (for GCC, e.g. 12, 13, 14; for LLVM, e.g. 15.0, 16.0. 17.0). That gives us a first degree of confidence that if the set of versions where the bugs appear overlaps but is not exactly the same, that this suggests two different bugs.

\emph{Commit screening/bisection}: For regressions, we know that there's a ``known working version of the compiler'' that doesn't exhibit the behavior. We exploit this to bisect and point to the first commit that introduced the violation. If the commit that introduces the new behavior is different for two different programs, then that very likely points to two different bugs. 

%% file: sections/results.tex
\section{Finding bugs in C / C++ compilers}

In this section we evaluate the effectiveness of our iterative mutation testing approach at finding missed optimization opportunities in C / C++ compilers.

\subsection{Experimental Setup}

To produce the experimental results in this section we ran each of the four differential testing strategies described previously for 8 hours. We use Llama 3.1~\citep{llama3} as the backing LLM, served on server-grade GPUs. We used two configurations of Llama 3.1: 70B parameters and 405B parameters.

In total we evaluate eight unique configurations of differential testing strategy and model for a total of 64 compute hours of testing. For each configuration we run a single threaded test loop, working through the iterative mutation testing process shown in Figure~\ref{fig:flow}. Code mutation episodes stop after 10 mutations if no violation has been found. The process then repeates until the time is used up.

\subsection{Results}

Table~\ref{tab:results} shows the results of the experiment. We evaluate each configuration using four metrics. The first, \emph{Total programs}, records the number of completed iterative tests. As can be seen, the 70B model produced significantly more programs than the 405B. \emph{Compilable} is the subset of Total programs which did not abort due to a compiler error. We see that the 405B model on average produces fewer compilation errors, but the difference is slight. The average over both models is 96.45\% successfully compilable code. \emph{Violations} is the number of times a program triggered the differential testing indicator of a suspicious compile, excluding those filtered out through the false positives detectors. In our experiments we found  that the larger 405B parameter LLM is the most effective model to find bugs, demonstrated here by a higher Violations rate. The tradeoff is that inference with the 405B model is more expensive than the smaller 70B model. That means that, for a fixed time period, the 70B model actually produced a higher total number of Violations, simply because the reduced efficiency was offset against far quicker inference, along many more programs to be generated in the same amount of time. Table~\ref{tab:inference-times} compares the inference times of the two models. It takes on average 2.92$\times$ longer to generate a code mutation using the 405B model than the 70B. During initial development we also experimented with the much smaller 8B parameter Llama model which offers yet faster inference, however, we found that this model so frequently corrupted the code with syntactic or semantic errors that it was impractical, and we abandoned this model configuration.

\begin{table*}[]
\centering
\caption{Results from 8 hours of automatic testing on each of the four modes of detecting suspicious compilations. The experiment is repeated using two configurations of the Llama 3.1 model: 70B parameters and 405B parameters. Results show that the larger 405B model has a higher rate of generating violations, but in absolute terms the slower inference means that for a given time budget, the smaller 70B model will discover a greater number of violations.}
\label{tab:results}
\scriptsize
\begin{tabular}{lcC{2cm}ccC{2cm}}
\toprule
\multicolumn{1}{c}{}                                                & \textbf{Model} & \textbf{Total programs} & \textbf{Compilable} & \textbf{Violations} & \textbf{Avg.\ steps (min / max)} \\ \hline
\multicolumn{1}{l}{\multirow{2}{*}{Dead code differential testing}} & 70B            & 5,598                   & 5,309 (94.84\%)     & 131 (2.34\%)        & 5.03 (2 / 10)                  \\
\multicolumn{1}{c}{}                                                & 405B           & 1,683                   & 1,598 (94.95\%)     & 40 (2.38\%)         & 4.17 (2 / 7)                   \\ \hline
\multirow{2}{*}{Optimization pipeline differential testing}                      & 70B            & 6,346                   & 5,905 (93.05\%)     & 145 (2.28\%)        & 4.89 (1 / 10)                  \\
                                                                    & 405B           & 1,933                   & 1,769 (91.52\%)     & 59 (3.05\%)         & 3.44 (2 / 7)                   \\ \hline
\multirow{2}{*}{Single-compiler differential testing}                         & 70B            & 6,305                   & 5,966 (94.62\%)     & 295 (4.68\%)        & 3.57 (1 / 10)                  \\
                                                                    & 405B           & 1,874                   & 1,760 (93.92\%)     & 124 (6.62\%)        & 2.67 (1 / 10)                  \\ \hline
\multirow{2}{*}{Multi-compiler differential testing}                          & 70B            & 10,008                  & 9,816 (98.08\%)     & 3,080 (30.78\%)     & 2.48 (1 / 10)                  \\
                                                                    & 405B           & 2,889                   & 2,786 (96.43\%)     & 1,035 (35.83\%)     & 2.30 (1 / 10)      \\            
\bottomrule
\end{tabular}
\end{table*}

\begin{table}
\centering
\caption{The minimum, mean, and max inference times of the two Llama 3.1 model configurations. Measurements aggregated from 32 hours of testing for both models.}
\label{tab:inference-times}
\begin{tabular}{cccc}
\toprule
\textbf{Model} & \textbf{Min} & \textbf{Mean} & \textbf{Max} \\
\midrule
70B           & 0.37s & 4.18s & 44.6s \\
405B           & 2.34s & 12.23s & 54.8s\\
\bottomrule
\end{tabular}
\end{table}

\subsection{Discovered bugs}

During development of our approach we discovered and reported 24 bugs in production compiles. The total compute time for our testing was about one week. In this section we present a representative sample of bugs found.

\begin{figure}
\setcaptiontype{lstlisting}
\begin{subfigure}{.95\columnwidth}
\begin{lstlisting}
long f() {
  long x = 0;
  while (x < 10) {
    if (x % 2 == 0) { x += 2; }
    else { x += 1; }
  }
  return x;
}
\end{lstlisting}
\caption{Correctly optimized code.}
\label{lst:bug1a}
\end{subfigure}
\begin{subfigure}{.95\columnwidth}
\begin{lstlisting}[mathescape=true]
long f() {
  long x = 0;
  while (x < 10) {
    if (x % 2 == 0) { x += 2; }
    else { x += 1; }

    $\textbf{\texttt{if ((x > 20) \&\& (x \% 5 == 0)) \{ x -= 5; \}}}$
    $\textbf{\texttt{if ((x < -5) \&\& (x \% 3 == 0)) \{ x += 3; \}}}$
  }
  return x;
}
\end{lstlisting}
\caption{Mutated code.}
\label{lst:bug1b}
\end{subfigure}
\caption{The addition of the two dead conditionals in (b) exposed a regression in GCC where Value Range Analysis fails to prove that the code is dead.}
\label{lst:bug1}
\end{figure}

\subsubsection{Bugs found by dead code differential testing}

We discovered and reported 5 bugs where compilers failed to identify and remove dead code (Section~\ref{sec:dead-code-difftest}).

\paragraph{GCC bug 116753} When the code in Listing~\ref{lst:bug1a} is compiled using the optimization pipeline \lstinline{-Os} using GCC trunk, the compiler is able to prove that the loop evaluates to a constant and simplifies the whole computation. But, if two dead conditions are added, as in Listing~\ref{lst:bug1b}, the optimizer is not able to optimize this code anymore. This is a regression in value range analysis, as this code used to work with an older version of GCC (12.4.0). The range analysis infrastructure has been reworked in GCC and the new pass can't eliminate the constraints anymore.

\paragraph{LLVM bug 112080} In Listing~\ref{lst:bug5}, the \lstinline{ConstraintElimination} pass in LLVM is not able to find that the inner loop is dead and remove it. After we reported the LLVM developer provided a patch that fixed the problem, and run this on the testsuite, showing positive results on real-world benchmarks. 

\begin{figure}
\setcaptiontype{lstlisting}
\begin{lstlisting}
long f() {
    long x = 0;
    while (x < 10) {
        while ((x > 20) && (x % 5 == 0)) {
            x -= 5;
        }
        if (x % 2 == 0) {
            x += 2;
        } else {
            x += 1;
        }
    }
    return x;
}
\end{lstlisting}
\caption{The \lstinline{ConstraintElimination} pass in LLVM is not able to find that the inner loop is dead and remove it. After reporting, this has been fixed upstream.}
\label{lst:bug5}
\end{figure}

\subsubsection{Bugs found by optimization pipeline differential testing}

We reported 4 bugs where optimization pipelines that are intended to reduce code size produce larger binaries than pipelines targeting runtime performance (Section~\ref{sec:pipeline-difftest}).

\paragraph{LLVM bug 111571} The code in Listing~\ref{lst:pipeline-bug} is fully optimized by LLVM trunk to a constant when the program is compiled with \lstinline{-O3}, but \lstinline{-Oz} fails to optimize it. This is because \lstinline{-O3} runs a loop pass that fully unrolls the loop, and later on, the dead store elimination pass finds out that all the assignments to the array are dead, removing this code. \lstinline{-Oz} doesn't run the unrolling pass and it's not able to prove this fact, leaving the code not optimized and blowing up the size. 

\begin{figure}
\setcaptiontype{lstlisting}
\begin{lstlisting}
void f() {
    int arr[5] = {1, 2, 3, 4, 5};
    for (int k = 0; k < 5; k++) {
        arr[k] *= 2;
    }
}
\end{lstlisting}
\caption{This code is optimized to a constant by LLVM when compiled using pipeline \lstinline{-O3}, but \lstinline{-Oz} fails to optimize it and generates a loop.}
\label{lst:pipeline-bug}
\end{figure}

\paragraph{GCC bug 117033} The code in Listing~\ref{lst:pipeline-bug2} shows that GCC trunk generates larger code at \lstinline{-Oz} than it generates at \lstinline{-Os}. In particular, GCC at \lstinline{-O3} can fold this code to a constant, while \lstinline{-Os} generates a loop. The \lstinline{-Os} pipeline is inconsistent about copying the loop header of the inner loop. That copy is critical to remove the outer loop, and subsequently, the Sparse Conditional Constant Propagation pass can't figure out the value using the dominator tree for the function.

\begin{figure}
\setcaptiontype{lstlisting}
\begin{lstlisting}
long f() {
    long x = 0;
    for (int i = 0; i < 5; ++i) {
        while (x < 37) {
            if (x % 4 == 0) {
                x += 4;
            }
        }
    }
    return x;
}
\end{lstlisting}
\caption{This code is optimized to a constant by GCC's \lstinline{-O3} pipeline but not by the \lstinline{-Os} pipeline.}
\label{lst:pipeline-bug2}
\end{figure}

\subsubsection{Bugs found by single-compiler differential testing}

We reported 12 code size regression bugs found by differential testing a compiler against older revisions of itself (Section~\ref{sec:single-compiler-difftest}).

\paragraph{GCC bug 117123} The code in Listing~\ref{lst:bug2} shows that GCC trunk generates larger code at \lstinline{-Os} than it does on GCC 13.3. The bisection points to \emph{scccopy}, a new optimization pass for copy propagation and PHI nodes elimination, developed in GCC 14.
Disabling this pass generates the same code on both versions of the compiler.
The GCC developers investigated this bug finding out that \textit{scccopy} just exposes a deficiency in the later partial redundancy elimination (PRE) pass, that misses an equivalence between PHI notes while doing value numbering.
The developers fixed this bug on trunk.

\begin{figure}
\setcaptiontype{lstlisting}
\begin{lstlisting}
int f(int a) {
  int arr[5];
  for (int i = 0; i < 5; i++) {
    if (i % 2 == 0 && a > 5 || i % 3 == 0 && a < -2) {
      arr[i] = a + i * 2;
    } else {
      arr[i] = a + i;
    }
  }
  return arr[2];
}
\end{lstlisting}
\caption{This code exposed a bug in the partial redundancy elimination pass of GCC, causing a code size regression between versions 13 and 14.}
\label{lst:bug2}
\end{figure}

\paragraph{GCC bug 117128} The code in Listing~\ref{lst:bug3} discovered a bug where GCC generates larger code at \lstinline{-Os} on trunk than GCC 14. The change that exposes this fact is a modification in the loop invariant code motion pass, but it's not the real culprit for the regression. In fact, trunk does some register allocation shrink-wrapping that doesn't happen on GCC 14. The GCC developers do concur that if shrink-wrapping in this case regresses size, it shouldn't be done at \lstinline{-Os}.

\begin{figure}
\setcaptiontype{lstlisting}
\begin{lstlisting}
struct Potato {   
  bool isMashed;   
};
void dont_be_here();   
int patatino_a;   
void patatino() {   
  if (patatino_a && patatino_a % 2 == 0 && patatino_a != 10)   
    ;   
  else {   
    Potato spud;   
    int spud_0 = patatino_a;   
    spud.isMashed = false;   
    for (int k = 0; patatino_a == 0 && spud_0 > 1000; k++)   
      for (int l = 0; l < 5; l++)   
        spud_0 += l * k;   
    for (; spud_0;)   
      dont_be_here();   
  }   
}
\end{lstlisting}
\caption{This code uncovered a regression in GCC 14 caused by the interaction of loop invariant code motion and shrink-wrapping passes.}
\label{lst:bug3}
\end{figure}

\subsubsection{Bugs found by multi-compiler differential testing}

We reported 3 missed optimization bugs found by comparing the binary sizes of code compiled using different compilers (Section~\ref{sec:multi-compiler-difftest}).

\paragraph{GCC bug 116868} The code in Listing~\ref{lst:bug4} bug shows that GCC can't prove that this function returns a vector with a single element, that's constant. This is because GCC can't safely prove that the allocation (via new) is ``sane'' (as defined by the C++ standard). Clang performs the expected optimization because it defaults to \lstinline{-fassume-sane-operator-new} as frontend flag, which asserts that a call to the operator \lstinline{new} has no side-effects beside the allocation, in particular that doesn't inspect or modify global memory. GCC just fixed this bug implementing the support. CSmith wouldn't be able to find this bug as it doesn't generate C++ code which contains calls to the standard library.

\begin{figure}
\setcaptiontype{lstlisting}
\begin{lstlisting}
#include <vector>
int sumVector() {
    const std::vector<int> vec = {1};
    int sum = 0;
    for (int i = 0; i < vec.size(); i++) {
        sum += vec[i];
    }
    return sum;
}
\end{lstlisting}
\caption{Clang will optimize this code to a constant. GCC lacks the required analysis to perform this optimization. The GCC developers are adding this feature to reach parity.}
\label{lst:bug4}
\end{figure}

\subsection{LLM code mutation failure cases}

On average 3.55\% of iterative test mutation sessions end because the LLM generated code that does not compile. In these cases we simply revert to the original seed code and restart mutation testing. Sometimes, the LLM will omit correctly compilable code but with unwanted or invalid semantics. The most common of these errors we observed is failure of the model to interpret the meaning of ``dead code''. We notice that sometimes the model interprets dead code as ``code that does nothing''. An example of this failure is shown in Listing~\ref{lst:not-dead-code}. In these cases, we use the dynamic instruction counts to filter out these invalid test cases, as described in Section~\ref{sec:false-positives}.


\begin{figure}
\setcaptiontype{lstlisting}
\begin{lstlisting}
int f(void) {
  int a = 0;
  for (;;) {
    a += 1;
    a -= 1;
  }
}
\end{lstlisting}
\caption{Incorrect LLM-mutated code in respond to a prompt instructing it to produce a dead loop. Here the loop has no side effects, but is not dead code since it would be reached during the normal flow of execution.}
\label{lst:not-dead-code}
\end{figure}

\section{Extending to other languages}

Because our technique uses an off-the-shelf LLM and automatic validation tools, it can easily be adapted to other languages by simply changing the prompt and the target compiler. To provide a preliminary evaluation of the extensibility of this approach we ported our 150 line Python script for testing C / C++ compilers to Rust and Swift. Adapting to each language is a trivial code change. We need only change the compiler invocation, and change the initial seed program as described in Section~\ref{sec:mutation}. We chose to target Rust and Swift languages as they are relatively young and so do not have the same infrastructure for random test case generation.

\subsection{Experimental Setup}

We modified the system to support Rust and Swift in turn by changing the language in the LLM prompt, and the compiler invocation commands. We then ran a single threaded instance of the testing loop for one hour each.

\subsection{Results}

Within a single hour of automated testing we discovered one bug in Swift and four bugs in Rust.

\paragraph{Rust bug 130421} The code in Listing~\ref{lst:bug6} produces a fully vectorized loop for something that can be simplified. This happens because the compiler relies on the loop to be fully unrolled to find out that the sum of the values is constant. Instead, the loop is just partially unrolled and vectorized, leading to larger code, and also slower computation of the result.

\begin{figure}
\setcaptiontype{lstlisting}
\begin{lstlisting}
pub fn foo() -> i64 {
    let mut result = 0;
    for i in 0..10 {
        for j in 0..10 {
            for k in 0..10 {
                if (i + j + k == 10) {
                    result += 1;
                }
            }
        }
    }
    return result
}    
\end{lstlisting}
\caption{The Rust compiler fails to fully simplify these nested loops because it relies on the loop to be fully unrolled to find out that the sum of the values is a constant.}
\label{lst:bug6}
\end{figure}

\paragraph{Swift bug 76535} The code in Listing~\ref{lst:bug7} uncovers a bug shows in Swift's compiler using \lstinline{-Osize} and \lstinline{-O} optimization pipelines. The compiler fails to establish that the nested while loop is dead and will never be reached. Removing the nested while loop does result in the loop being simplified. This could be caught in Swift if one of the SIL passes (the Swift intermediate language) would implement a range analysis mechanism to remove the dead computations.

\begin{figure}
\setcaptiontype{lstlisting}
\begin{lstlisting}
func f() -> Int {
  var numbers = [1,2,3,4,5]
  var sum = 0
  for number in numbers {
    if (number > 2 && number < 5) ||
       (number == 1 || number == 3) || 
       (number % 2 == 0 && number > 1) || 
       (number % 3 == 0 && number < 4) || 
       (number * 2 == 6 && number + 1 == 4) || 
       (number - 1 == 2 && number / 2 == 1) {
      sum += number
      while (number + 1 == 5 && number * 2 == 9) {
      }
    }
  }
  return sum
}
\end{lstlisting}
\caption{The Swift compiler fails to establish that the nested while loop is dead and will never be reached.}
\label{lst:bug7}
\end{figure}

\paragraph{Rust bug 132888} The code in Listing~\ref{lst:bug8} when compiled with the Rust nightly compiler is larger than when it is compiled with Rust 1.81.0. The backend of the compiler decides to emit a different sequence of instructions that increase codesize overall for both \lstinline{x86-64} and \lstinline{aarch64}.

\begin{figure}
\setcaptiontype{lstlisting}
\begin{lstlisting}
#![no_main]
#[no_mangle]
pub struct Point {
    x: i32,
    y: i32,
}

#[no_mangle]
pub fn f(a: Point) -> i32 { 
    if a.x > 0 && a.y < 0 || a.x < 0 && a.y > 0 {
        a.x * a.y
    } else {
        a.x + a.y
    }
}
\end{lstlisting}
\caption{The size of this Rust code regressed by 50\% between 1.81.0 and trunk.}
\label{lst:bug8}
\end{figure}

\paragraph{Rust bug 132890} The code in Listing~\ref{lst:bug9} shows a simple array iteration, and results in a larger binary when compiled with Rust nightly than when compiled with Rust 1.73.0. This is another example of the compiler backend lowering a suboptimal sequence of instructions. The model in this example the LLM generates code with non trivial API usage (e.g. \lstinline{chain()}, \lstinline{iter()}).

\begin{figure}
\setcaptiontype{lstlisting}
\begin{lstlisting}
#![no_main]
#[no_mangle]
pub fn f(a: i32) -> i32 { a + a }

#[no_mangle]
pub fn g(a: [i32; 5]) -> i32 {
    let mut sum = 0;
    let arr = [1, 2, 3, 4, 5];
    for i in a.iter().chain(arr.iter()) {
        sum += i;
    }
    sum
}
\end{lstlisting}
\caption{The Rust compiler emits a suboptimal sequence of instructions in the backend for this example.}
\label{lst:bug9}
\end{figure}

%% file: sections/related-work.tex
\section{Related Work}

Automatic test case generation for compilers is a well established technique for compiler validation and has been surveyed in~\citep{chen2020survey,tang2020compiler}. Typically random program generators are in themselves complex pieces of code. For example, CSmith~\citep{csmith} is over 40,000 lines of handwritten C++. We take a different approach. Inspired by mutation-based approaches such as equivalence modulo inputs testing~\citep{emi}, we adopt a process of mutation testing, but starting from a trivial seed program, and using an LLM as the engine for rewriting code. Our work differs from previous mutation-based approaches in that it does not require existing libraries to begin mutation, such as the \lstinline{libc} functions from FreeBSD used in~\citep{boosting}.

Prior works using machine learning to synthesize compiler test cases include~\citep{fuzz4all,deepsmith,deepfuzz,learn-and-fuzz}. Of those, the closest to our work is Fuzz4all~\citep{fuzz4all}, a recent publication that employs Large Language Models to generate novel compiler tests. Their approach first ingests documentation and example code and uses it to synthesize new test cases to validate functional correctness of the compiler. Our work differs in that it targets missed optimization opportunities rather than functional correctness, requires no documentation as input, and incrementally builds complexity by mutating a trivial starting seed, rather than synthesizing novel test cases from scratch.

While much attention has been placed on validating the functional correctness of compilers, relatively little attention has been targeted on identifying missed optimizations. Two recent works that attempt to identify missed optimizations are~\citep{theodoridis2022finding} and~\citep{liu2023exploring}. In~\citep{theodoridis2022finding}, missed optimizations are identified by instrumenting the basic blocks of CSmith-generated code using ``dead code markers''. In~\citep{liu2023exploring}, a C program is compiled using both x86 and WebAssembly compilers to identify missed optimizations in WebAssembly.
Compared to both these works, our approach is not language specific, requires no instrumentation of programs, and is the first work to use machine learning to generate code rather than handcrafted rules.

%% file: sections/conclusions.tex
\section{Conclusions}

We describe our experience using Large Language Models to identify missed code size optimization opportunities in compilers. We start with C / C++ and found that by orchestrating an off-the-shelf LLM and existing software validation tools, an effective yet remarkably simple approach could be taken, yielding 24 bugs in production compilers. While our initial results are promising, we are just scratching the surface. In future work we will extend the differential testing methodologies to detect missing runtime performance optimizations, explore prompt engineering approaches to improve the effectiveness of the approach further, and mutate larger seed codes. We hope our initial results to inspire interest in this exciting research direction.